\documentclass[11pt]{article}
\usepackage{amsmath}
\usepackage{amssymb}
\usepackage[dvips]{graphicx}
\usepackage[usenames]{color}
\usepackage[doublespacing]{setspace}
\usepackage{setspace}
\usepackage[authoryear]{natbib}

\begin{document}
\doublespacing

\title{ Regulative Differentiation as Bifurcation of Interacting Cell Population}
\date{}
\maketitle
\vspace{-10mm}
\noindent
Akihiko Nakajima$^{1}$, Kunihiko Kaneko$^{1,2}$\vspace{3mm}\\
$^{1}$Department of Basic Science, University of Tokyo, 3-8-1 Komaba, Meguro-ku,
Tokyo 153-8902, Japan\\
$^{2}$ERATO Complex Systems Biology Project, JST, Japan\vspace{3mm}\\
\noindent
Corresponding author: Akihiko Nakajima\\
E-mail address: nakajima@complex.u-tokyo.ac.jp\\
Tel./fax: +81 3 5454 6732.\\

In multicellular organisms, several cell states coexist.
For determining each cell type, cell-cell interactions are often essential,
in addition to intracellular gene expression dynamics.  Based on dynamical systems theory,
we propose a mechanism for cell differentiation with regulation of populations of each cell type
by taking simple cell models with  gene expression dynamics.
By incorporating several interaction kinetics, we found that the cell models
with a single intracellular positive-feedback loop exhibit a cell fate switching,
with a change in the total number of cells.  The number of a given cell type or the
population ratio of each cell type is preserved against the change in the total number of cells,
depending on the form of cell-cell interaction. The differentiation is a result of bifurcation of
cell states via the intercellular interactions, while the population regulation is explained
by self-consistent determination of the bifurcation parameter through cell-cell interactions.
The relevance of this mechanism to development and differentiation in several multicellular systems is discussed.


 \section{Introduction}
\indent\indent
 Complex gene regulatory or protein networks are responsible for determining cellular
behaviors.  The function of such networks has recently been discussed in the light of
specific network structures called network motifs~\citep{Shen-Orr02,Milo02,Milo04}.
Besides such motifs, several simple network modules are also considered to operate
to give specific dynamical properties such as bistability, adaptation, or oscillatory behavior~\citep{Ferrell98,Sha03,Tyson03}.
Recent experimental results also suggest that
such modules provide a basis for cell differentiation, as studied in
competence state in {\it Bacillus subtilis}~\citep{Suel06,Maamar07}.
   \\

   \indent
   In multicellular organisms, several cell states coexist.  Morphogenesis
with differentiation into distinct cell types, however, is not an event of independent
single-cellular dynamics, but occurs as a result of an ensemble of interacting cells.
For determining each cell type, cell-cell interactions are often essential
besides intra-cellular dynamics by functional modules at a single cell level.
In fact, gene regulatory networks responsible for the early developmental process
or the cell specification process of several kinds of  organisms
include many intercellular interactions~\citep{Ben-Tabou07,Davidson02,Imai06,Loose04,Swiers06}.  The importance of
cell-cell interactions to robust developmental processes is discussed
as the community effect~\citep{Gurdon93} and differentiation from
equivalent groups of cells~\citep{Greenwald92}.\\

When considering the development of a multi-cellular organism, not only a set of cell types, 
but also the number distribution of each of the cell types, has to be suitably determined and robust
against perturbations during the course of development.  
The proportion of the body plan in planarian and in the slug of {\it Dictyostelium discoideum} is 
preserved over a wide range of body sizes~\citep{Oviedo03,Rafols01}. In the {\it D. discoideum} slug, 
the number ratio of two cell types  is kept almost constant.  In the hematopoietic system of mammals
approximately ten different cell types are generated from a hematopoietic stem cell,
and their growth and differentiation are regulated to keep the number distribution of each cell to achieve homeostasis of the hematopoietic system.  In this case, in addition to the proportion regulation, 
the absolute size of stem cells is also important because all the hematopoietic cells will ultimately die out without their existence.
Indeed, regulation of the numbers of each cell type is rather common in multicellular organisms.
As the distribution of each cell type is a property of an ensemble of cells,
cell-cell interactions should be essential for such regulation.\\

   \indent
   There are several theoretical studies discussing the importance of cell-cell
   interactions. By considering an ensemble of cells with intra-cellular genetic (or chemical) networks 
   and intercellular interactions, synchronization of oscillation~\citep{Garcia-Ojalvo04,McMillen02}
   or dynamical clusterings~\citep{Kaneko94, Mizuguchi95, Kaneko97, Furusawa98, Ullner07, Koseska07} are observed.
   Cell states distinguishable from those of a single-cellular dynamics are generated, providing a basis for functional
   differentiation for multicellularity.
   The preservation of the proportion of different cell types
   is realized by taking advantage of Turing instability~\citep{Mizuguchi95}, while the robustness in the number distribution of
   different cell types is discovered in reaction network models~\citep{Kaneko94, Furusawa98, Kaneko99}.  Nevertheless,
   regulatory mechanisms for cell type populations are not elucidated in terms of dynamical
   systems because of the high dimensionality of the models.\\

   \indent
   In the present paper, we propose
   a regulatory mechanism of cell differentiation based on dynamical systems theory 
   by taking simple cell models with biological gene regulation dynamics.
   Specifically, we study how cell states are differentiated with the change in the total
   cell number following  cell-cell interactions.
   By incorporating different interaction kinetics, we show  
   how simple functional modules generate specific cellular behaviors
   such as a cell fate switching, 
   size regulation of each cell type,
   and preservation of the number ratio of each cell type.\\

   \indent
   The present paper is organized as follows.
   In section 2, we introduce an interacting multicellular model
   which is further analysed in the present paper.
   Each cell has a simple functional module of genes,
   and its expression dynamics is modulated 
   by the interactions with other cells.
   In sections 3 to 5,
   we consider several different intercellular interactions, respectively.
   Although possible cell states are generated by an intracellular functional module,
   selection of one of these possible states or
   establishment of specific number distributions of cell states 
   is realized depending on the manner of intercellular interactions.
   In section 6, we extend our theoretical scheme to discuss the
   distribution of cell types in cell differentiation models 
   studied so far~\citep{Kaneko97,Kaneko99,Furusawa98,Furusawa01}.
    Although these models have a complex intra-cellular reaction network,
    we show that the same logic can be applied to explain the cell differentiation 
   observed in these models.   In section 7, we summarize our results and discuss their biological
    relevance and future directions.

     \section{Framework of the Model}
     \indent\indent
     Here we introduce a basic model of interacting cells with
     intracellular gene 
     expression dynamics. Consider $N$ cells with identical genes
     which interact through a common medium.  The internal state of $i$-th cell
     is represented by the expression pattern of $m$ genes,
     as $\vec{u_i}=(u_i^1,\ldots,u_i^m)^T$. 
     The medium under which cells are placed is represented by concentrations
     of $n$ diffuse signals given by
     $\vec{v}=(v_1,\ldots,v_n)^T$. As the simplest case, we
     discard the spatial configuration of cells so that each cell interacts
     with all the other cells via common signal chemicals
     $\vec{v}$. Each intracellular gene expression dynamics is modulated
     by these signal molecules, which give interactions with other
     cells. \\
     
     For the sake of simplicity, we mostly examine the dynamics of single gene
     expression,  in which the state of the $i$-th cell is expressed by only one
     variable, $u_i$, and the intercellular interaction is mediated by only
     one global diffusive signal, $v$.  By using the standard kinetics of gene
     expression, $\{u_i\}$ and $v$ are chosen to obey the following equation,

     \begin{align}
      \frac{du_i(t)}{dt} &=
      f(u_i,v) = 
      \frac{1}{\tau} \left(
      \frac{u^{\alpha}_i(t)}{K^{\alpha}_u+v^{\alpha}(t)+u^{\alpha}_i(t)}-u_i(t)+A_u
      \right) &\mbox{for }i=1,\ldots,N, \label{a1}\\
      \frac{dv(t)}{dt} &= {\rm g}(u_1,\dots,u_N,v).\label{a2}
     \end{align}
     Here  gene $u_i$ activates its own expression by feedback,
     while the signal $v$ has an inhibitory effect on the expression of the
     gene $u_i$.
     Generally, the signal $v$ is released by each cell depending 
     on its gene expression level and the signal abundances at that moment. 
     We adopt Hill-type kinetics for self activation of the gene
     $u_i$. The parameter $\alpha$ denotes the Hill coefficient, i.e., the
     cooperativity of its kinetics, while $K_{u}$ is the threshold for the
     activation of gene $u_i$, and $A_{u}$ is the activation rate of $u_{i}$
     by other molecules in the cell. The parameter $\tau$ is a time constant
     of the expression dynamics of $u_i$ normalized by that of the signal
     $v$.  
     In the present paper, we consider that the timescale of $u_{i}$ is much
     slower than that of $v$, so that only fixed point solutions are
     realized. The assumption on the time scale is biologically reasonable because the
     gene expression process requires a much longer time than simple catalytic reactions. 
     For numerical simulations, we use the following parameter values;
     $K_{u}=0.1$, $A_{u}=0.04$, ${\alpha}=2.0$, ${\tau}=10.0$.
     Note that the following results are qualitatively invariant
     as long as the Hill-coefficient $\alpha$ is larger
     than unity.\\

\begin{figure}[htb]
     \begin{center}
      \includegraphics[width=12cm]{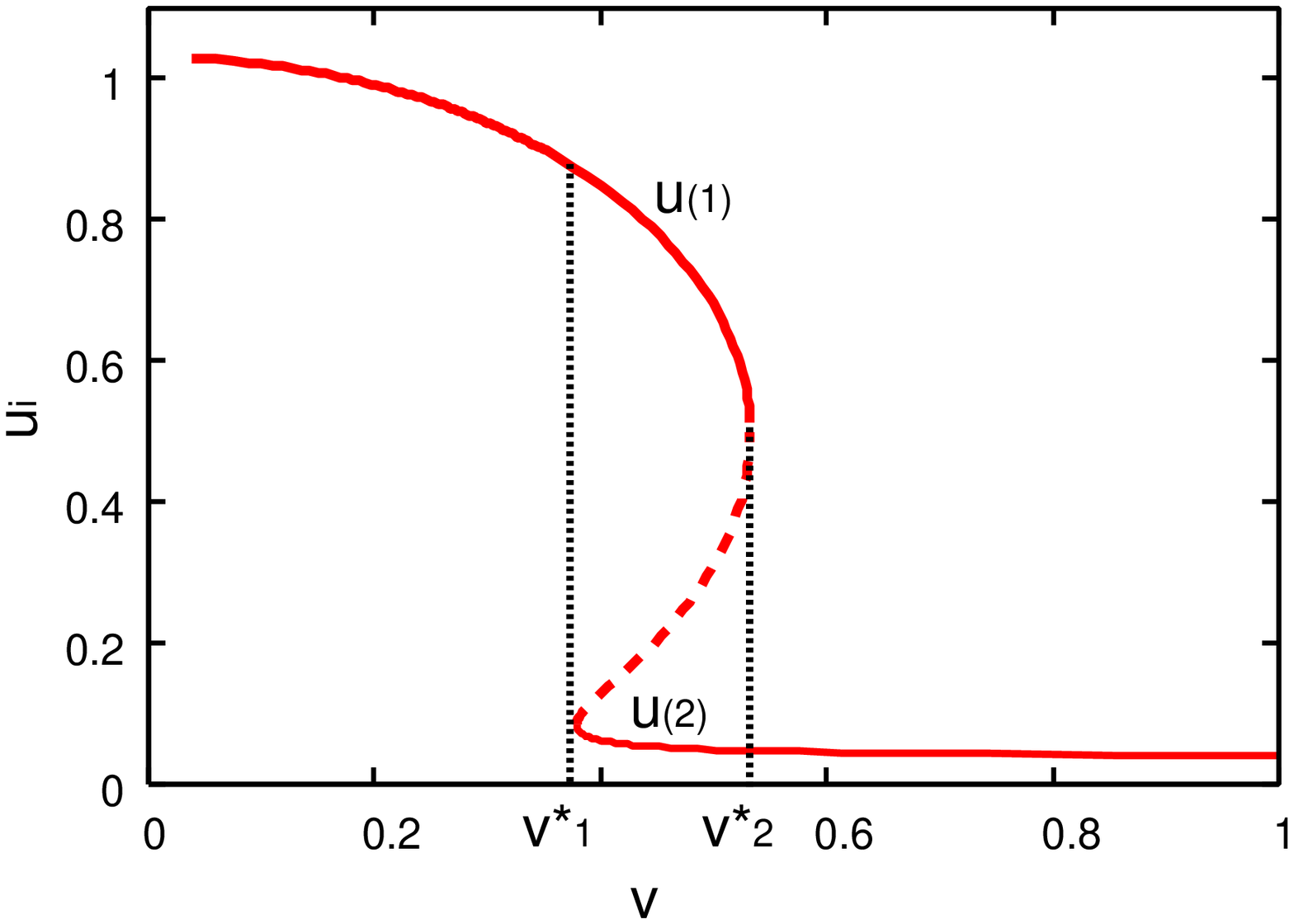}
      \caption{}
      \label{fig1}
    \end{center}
  \end{figure}

     \indent
     Before studying the dynamics of a population of interacting cells,
     we first survey the single intracellular dynamics Eq. (\ref{a1})
     with $v$ given as a constant control parameter. As is shown
     straightforwardly, the equation has a fixed point solution which
     exhibits two saddle-node bifurcations with the change in $v$
     (Fig. \ref{fig1}). We denote these bifurcation points as
     $v=v^{*}_1$ and $v=v^{*}_2$, and call the upper branch of the
     stable state as $u_{(1)}$ (or cell state 1) that is stable at $v
     \leq v^{*}_2$, and the other lower branch as $u_{(2)}$ (or cell
     state 2) that is stable at $v \geq v^{*}_1$.
     In the parameter region $v_1^* <v< v_2^*$, the bistability of
     $u_{(1)}$ and $u_{(2)}$ is sustained.\\

     \indent
     As shown in Fig. \ref{fig1}, the only possible stationary states
     of each cell are $u_i=u_{(1)}$ or $u_i=u_{(2)}$. Depending on the
     value of $v$ and also on the initial condition of $u_i$, each of the two
     solutions are selected. The question we address is as follows: 
     how are these states selected and what determines a possible range 
     in the number distribution of the two states
     when intercellular interactions through $v$ are taken into account. 
     In the following sections, we analyze three models with different types of
     the function ${\rm g}(u_1,\dots,u_N,v)$ 
     to study how the differences in the kinetics of $v$ 
     lead to different types of regulation in the number distribution of cell types.

     \section{Model I: Cell Fate Determination by Total Cell Number}
     \indent\indent
     As a first example of interacting cells, we adopt a model in which
     each cell simply emits the signal $v$ with the same
     rate. The kinetics of $v$ obeys the following equation,
     \begin{align}
      \intertext{{\bf model I}}
      \frac{dv(t)}{dt} &= {\rm g_1}(u_1,\dots,u_N,v)=
      \sum_{i=1}^{N}{c_i}-v(t) = c_{1}N-v(t),\label{a3}
     \end{align}
       while the kinetics of $\{u_i\}$ obey Eq. (\ref{a1}). 
       We are interested in the behavior of the stationary state as a
       function of the total cell number $N$. The stationary state
       solution of an ensemble of cells is generally obtained by the
       following procedure. First, we regard the signal $v$ as a fixed
       parameter, not a variable, and obtain the solution $u_{i}$ as a
       function of $v$, as already described in the previous
       section. Next, we write down $v$ as a function of $N$ and
       $\{u_{i}\}$ so that the self-consistent solution of the coupled
       equation is obtained, from which we analyze the
       dependence of the solution on the total cell number.\\

\begin{figure}[htb]
     \begin{center}
    \includegraphics[width=12cm]{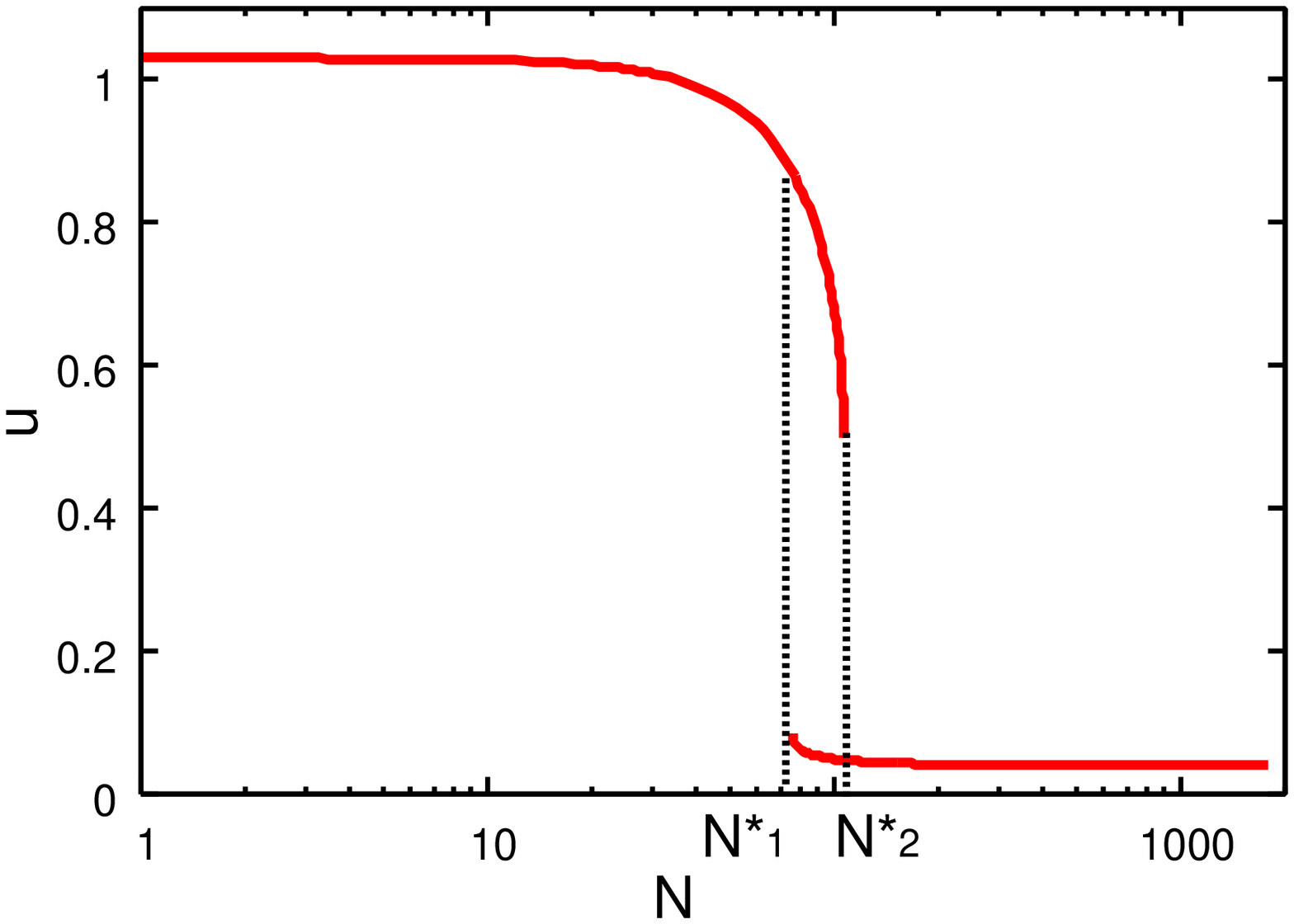}
   \caption{}
    \label{fig2}
    \end{center}
  \end{figure}

\begin{figure}[htb]
     \begin{center}
     \includegraphics[width=12cm]{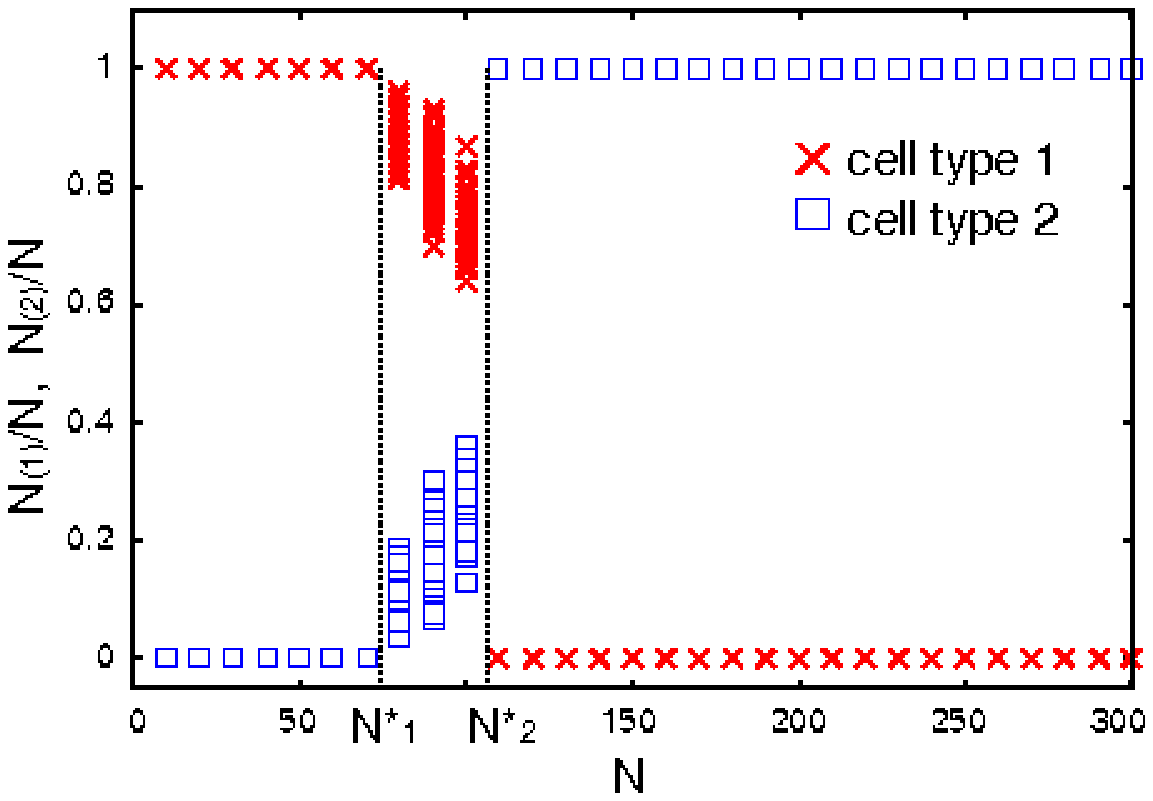}
    \caption{}
    \label{fig3}
    \end{center}
  \end{figure}

       \indent
       The stationary state is simply obtained by $du_i/dt=0$
       and $dv/dt=0$. In the present case, the solution $v$ is
       independent of $\{u_i\}$, and depends only on $N$, which leads to
       \begin{align}
	f(u_i,v)&=0,\hspace{0.5cm}v=c_{1}N.
       \end{align}
       The solution curve $f(u_i,v(N))=0$ is shown in
       Fig. \ref{fig2}, and the numerical result of the ratio of the
       number of each cell type to the total cell number is shown in
       Fig. \ref{fig3}. 
       Here we define a single-cluster of an ensemble of cells as a state
       in which all the cells take the same stationary states, i.e.,
       \begin{align}
	u_{i}=u_{(k)}~~(k=1,~\mbox{or}~2)\hspace{1cm}\mbox{for }i=1,2,\ldots,N,
       \end{align}
       and a two-cluster state as that in which two cell types with $u=u_{(1)}$ and
       $u=u_{(2)}$ coexist, so that
	  \begin{align}
	   u_{i} &=
	   \begin{cases}
	    u_{(1)} &\mbox{for}~i=1,\ldots,N_{(1)},\\
	    u_{(2)} &\mbox{for}~i=N_{(1)}+1, \ldots, N_{(1)}+N_{(2)}(=N).\\
	   \end{cases}
	  \end{align}
	  Here $N_{(1)}$ and $N_{(2)}$ denote the number of the cells with
	  $u=u_{(1)}$ and $u=u_{(2)}$, respectively.\\

          \indent
	  When the cell number $N$ is lower than a
	  threshold $N^{*}_{1}$ ($= v_1^{*}/c_1$), the single-cluster state of
	  $u_{(1)}$ is realized, while
	  for $N$ larger than a threshold $N^{*}_{2}$ ($= v_2^{*}/c_1$), 
	  the single-cluster state of $u_{(2)}$ is realized,
	  irrespectively of the initial cell state.
	  Only within the range of $N^{*}_{1} \leq N \leq N^{*}_{2}$ 
	  are two-cluster states of $u_{(1)}$ and $u_{(2)}$ possible,
	  where any population ratio of the cell types with $u_{(1)}$ to
	  $u_{(2)}$ can be realized depending on the initial condition.
	  Cell types switch between $u_{(1)}$ and $u_{(2)}$
	  simply by the total cell number, and the signal $v$ works as a 
	  population size detector.

    \section{Model I\hspace{-.1em}I: Diversification from Single State,
 and Size Regulation of Specific Cell Type}
  \indent\indent
  Next, we consider the case in which the signal induction depends on the expression
  level of $u_i$. We will show that the cells are
  differentiated into two types over a wide range of the total cell number
  $N$, and that the number of type 1 cells remains at a same level herein.\\

  \indent
  The kinetics of the signal $v$ in model \mbox{I\hspace{-0.1em}I} is represented as follows,
   \begin{align}
    \intertext{{\bf model I\hspace{-0.1em}I}}
    \frac{dv(t)}{dt} &=
    {\rm g_2}(u_1,\dots,u_N,v)=
    c_{2}\sum_{i=1}^{N}{\frac{u^{\beta}_i(t)}
    {K_v^{\beta}+u^{\beta}_i(t)}}-v(t).\label{a4}
   \end{align}
   We here adopt Hill-type kinetics for the induction of the signal
   $v$ by $u_i$, where $\beta$ is the Hill coefficient, representing the
   cooperativity in the induction, and $K_{v}$ denotes the threshold
   value for the signal induction. The parameter $c_{2}$ gives the
   release rate of $v$ from each cell.\\

\begin{figure}[htb]
     \begin{center}
      \includegraphics[width=12cm]{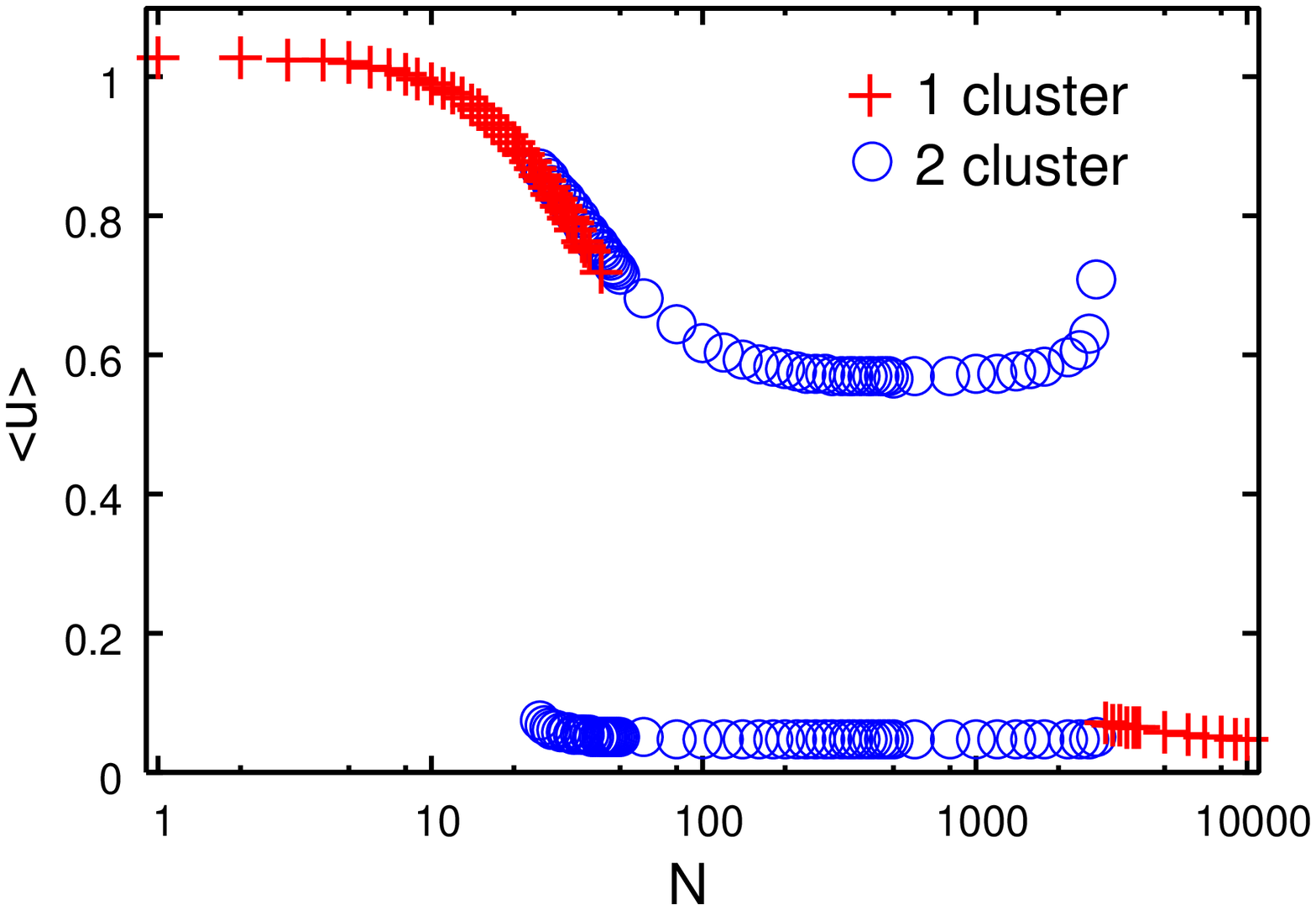}
      \caption{}
      \label{fig4}
    \end{center}
  \end{figure}

   \indent
   Dependence of the stationary states on the total cell
   number is shown in Fig. \ref{fig4}.
   For a small $N$, all the cells always fall on a single-cluster state of $u_{(1)}$.
   As $N$ gets larger, the bifurcation to a two-cluster state occurs, where
   the cells take either $u_{(1)}$ or $u_{(2)}$.
   Here, the single-cluster state of $u_{(1)}$ ($u_{(2)}$) is realized only
   at a small (large) number of cells, respectively, so that there is a
   gap in the total number of cells between the two single-cluster states. 
   The two-cluster state exists within this gap.\\

\begin{figure}[htb]
     \begin{center}
      \includegraphics[width=12cm]{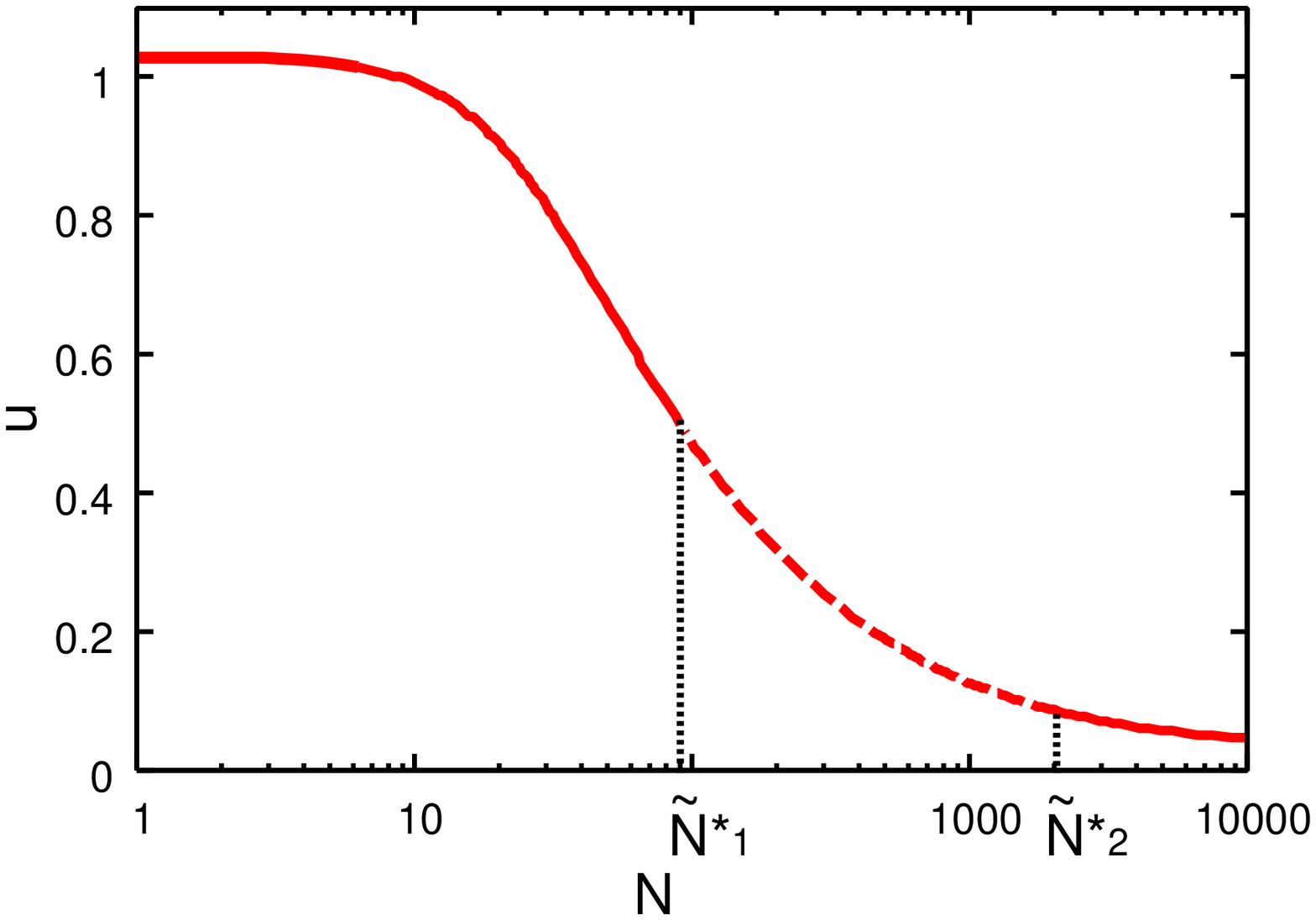}
      \caption{}
      \label{fig5}
    \end{center}
  \end{figure}

   \indent
   To understand the observed dependence of the clustering behavior on
   the cell number,
   we first consider the stability of a single-cluster state.
   From $du_i/dt=0$, $dv/dt=0$, and $u_i=u_{(k)}$ 
   ($k=1,~\mbox{or}~2$) for $i=1,\ldots,N$, we get
   \begin{align}
    f(u_{(k)},v)&=0,\hspace{0.5cm}v=c_{2}N 
    \frac{u^{\beta}_{(k)}}{ K_v^{\beta} + u^{\beta}_{(k)}}.
    \end{align}
   By solving the above equations self-consistently, 
   the solution curve of $u$ is obtained 
   as a function of the total cell number $N$
   (Fig. \ref{fig5}). 
   For $N < \tilde{N}_{1}^{*}$,
   a single-cluster state of $u_{(1)}$ is always stable.
   When the cell number increases beyond $\tilde{N}_{1}^{*}$, this
   single-cluster state becomes unstable, while for much larger
   $N$ such that $N>\tilde{N}^{*}_{2}$, the single-cluster state becomes
   stable again, where the cell state is  $u_{(2)}$  (Fig. \ref{fig5}) .
   The threshold $\tilde{N}_{1}^{*}$ and $\tilde{N}_{2}^{*}$ are given by 
   $\tilde{N}^{*}_{1} = v^{*}_{2}(K_v^{\beta} +
   u^{\beta}_{(1)}(v^{*}_{2}))/(c_{2}u^{\beta}_{(1)}(v^{*}_{2}))$ and 
   $\tilde{N}^{*}_{2} = v^{*}_{1}(K_v^{\beta} +
   u^{\beta}_{(2)}(v^{*}_{1}))/(c_{2}u^{\beta}_{(2)}(v^{*}_{1}))$, respectively.\\

   \indent
   Next, consider the condition for the existence of a two-cluster state.
   Because the stability of a cell state is determined
   by the amount of $v$, the condition for the existence of 
   a two-cluster state is given by $v_{1}^{*}<v<v_{2}^{*}$.
   Accordingly, considering $v$ as a function of $N_{(1)}$ and $N$,
   a two-cluster state is possible if $N_{(1)}$ 
   satisfies $v_{1}^{*}<v(N_{(1)},N)<v_{2}^{*}$.
   Note that $v$ satisfies ${\partial v(N_{(1)},N)}/{\partial N_{(1)}} >0$.   
   Thus, the range of the cell number $N$ in which a two-cluster state
   exists is given by 
   $N^{*}_{1}< N <N^{*}_{2}$, where 
   $N^{*}_{1} = v^{*}_{1}(K_v^{\beta} +
   u^{\beta}_{(1)}(v^{*}_{1}))/(c_{2}u^{\beta}_{(1)}(v^{*}_{1}))$ and 
   $N^{*}_{2} = v^{*}_{2}(K_v^{\beta} +
   u^{\beta}_{(2)}(v^{*}_{2}))/(c_{2}u^{\beta}_{(2)}(v^{*}_{2}))$, 
   respectively.
   These threshold sizes satisfy $N_{1}^{*}<\tilde{N}_{1}^{*}$ and
   $\tilde{N}_{2}^{*} < N_{2}^{*}$, so that only two-cluster states are
   stable for N satisfying $\tilde{N}_{1}^{*}< N < \tilde{N}_{2}^{*}$.\\

\begin{figure}[htb]
     \begin{center}
      \includegraphics[width=12cm]
      {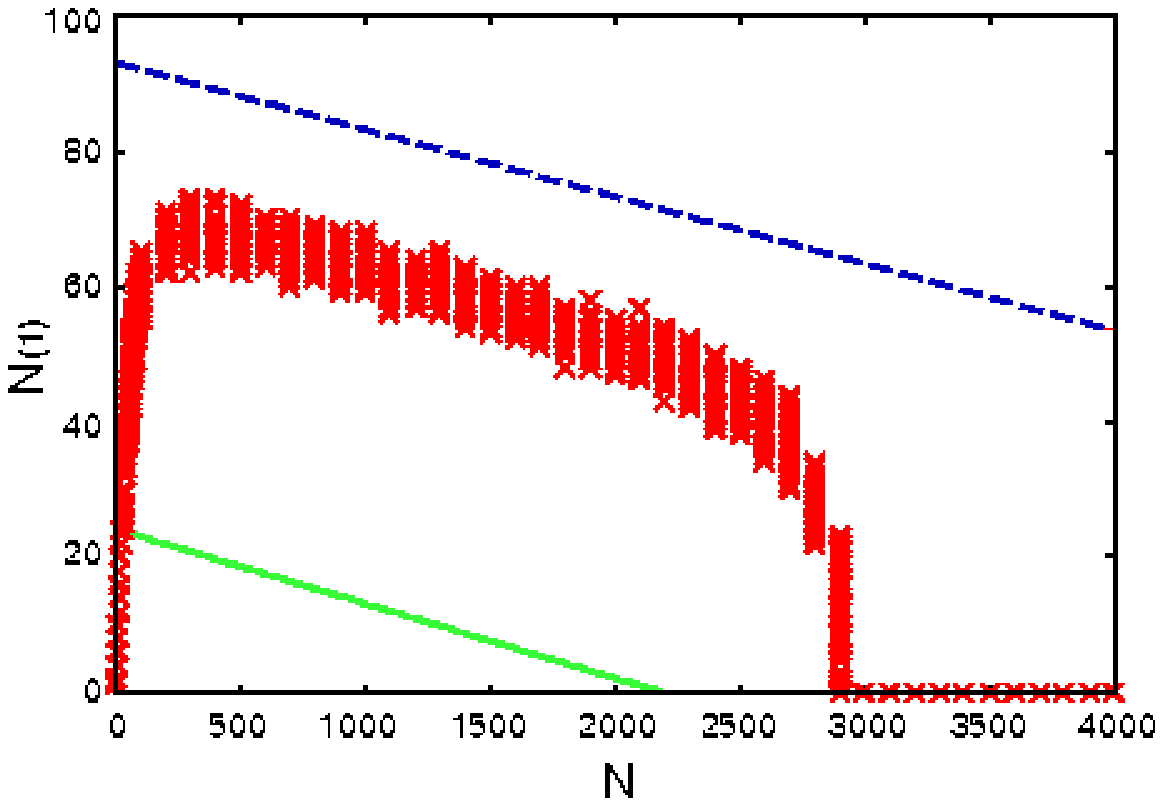}
      \caption{}
      \label{fig6}
    \end{center}
  \end{figure}

   \indent
   Because the number of each cell type in these two-cluster states
   has to satisfy the above condition, the range of possible numbers of
    two cell types is limited, depending on the total number of cells.
    The number of cell type 1 ($N_{(1)}$) from a variety of initial
    conditions is plotted as a function of $N$ in Fig. \ref{fig6}.  As $N$ is increased beyond
    $N_{1}^*$, $N_{(1)}$ decreases linearly with $N$, with a rather small slope,
    over a wide range of $N$, up to $N_{2}^*$.  Within this range the value
    of $N_1$ does not change so much.\\
    
    \indent
   To understand this behavior we obtain the dependency of $N_{(1)}$ on 
   $v$ and $N$. In a two-cluster state $(N_{(1)},N_{(2)})$,  $v$ is expressed by 
   the contribution from the cell types 1 and 2. 
       Thus, $N_{(1)}$ is written as 
    \begin{align}
     N_{(1)}(N,v) &= -A(v)N+B(v), \label{a10}
    \end{align}
    \begin{align}
     A(v)
     &= \frac{u_{(2)}^{\beta}(v)/(K_{v}^{\beta}+u_{(2)}^{\beta}(v))}
     { u_{(1)}^{\beta}(v)/(K_{v}^{\beta}+u_{(1)}^{\beta}(v))
     -u_{(2)}^{\beta}(v)/(K_{v}^{\beta}+u_{(2)}^{\beta}(v)) },\label{eq10}\\
     B(v)
     &= \frac{v}{c_2 \{ u_{(1)}^{\beta}(v)/(K_{v}^{\beta}+u_{(1)}^{\beta}(v))
     -u_{(2)}^{\beta}(v)/(K_{v}^{\beta}+u_{(2)}^{\beta}(v))\} }.
     \end{align}
     Here, we note that
     $u_{(1)}$ and $u_{(2)}$ are determined self-consistently as functions of
     $v$,
     and that $A(v)>0$ and $B(v)>0$. 
     For the existence of a two-cluster state, $v$
     has to satisfy $v^{*}_1< v < v^{*}_2$, that is, $N_{(1)}(N,v^{*}_1)
     < N_{(1)}(N,v) < N_{(1)}(N,v^{*}_2)$ for each $N$.  
     By inserting Eq. (\ref{a10}) into this expression, it is shown that 
     $N_{(1)}(N, v_1^*)$ and $N_{(1)}(N, v_2^*)$, i.e., the lower and upper
     bounds of $N_{(1)}$, decay linearly with $N$, with the slope
     of $A(v_1^*)$ and $A(v_2^*)$.
     In fact, a linear decrease in $N_{(1)}$ with the increase in $N$ is clearly
     discernible in Fig. \ref{fig6}.\\

     \indent
     Next, we evaluate the value of the slope $A(v)$.
     Eq. (\ref{a10}) is written as 
     $A(v)= \{ {(u_{(2)}/K_v)}^{\beta}+{(u_{(2)}/u_{(1)})}^{\beta} \}
     /\{ 1- {(u_{(2)}/u_{(1)})}^{\beta} \}$.
     If $u_{(2)} \ll u_{(1)}$ and $u_{(2)} \ll K_{v}$ are satisfied, 
     that is the case for the parameters used in Fig. \ref{fig6},
     $A(v)$ is much smaller than unity. 
     As a result, the decrease in $N_{(1)}$ with $N$ is slow, 
     and $N_{(1)}$ is sustained at a same level over a
     wide range of $N$, satisfying $N_{(1)}(v^{*}_1) < N_{(1)}(v)
     < N_{(1)}(v^{*}_2)$ (Fig. \ref{fig6}).\\

\begin{figure}[htb]
     \begin{center}
      \includegraphics[width=12cm]{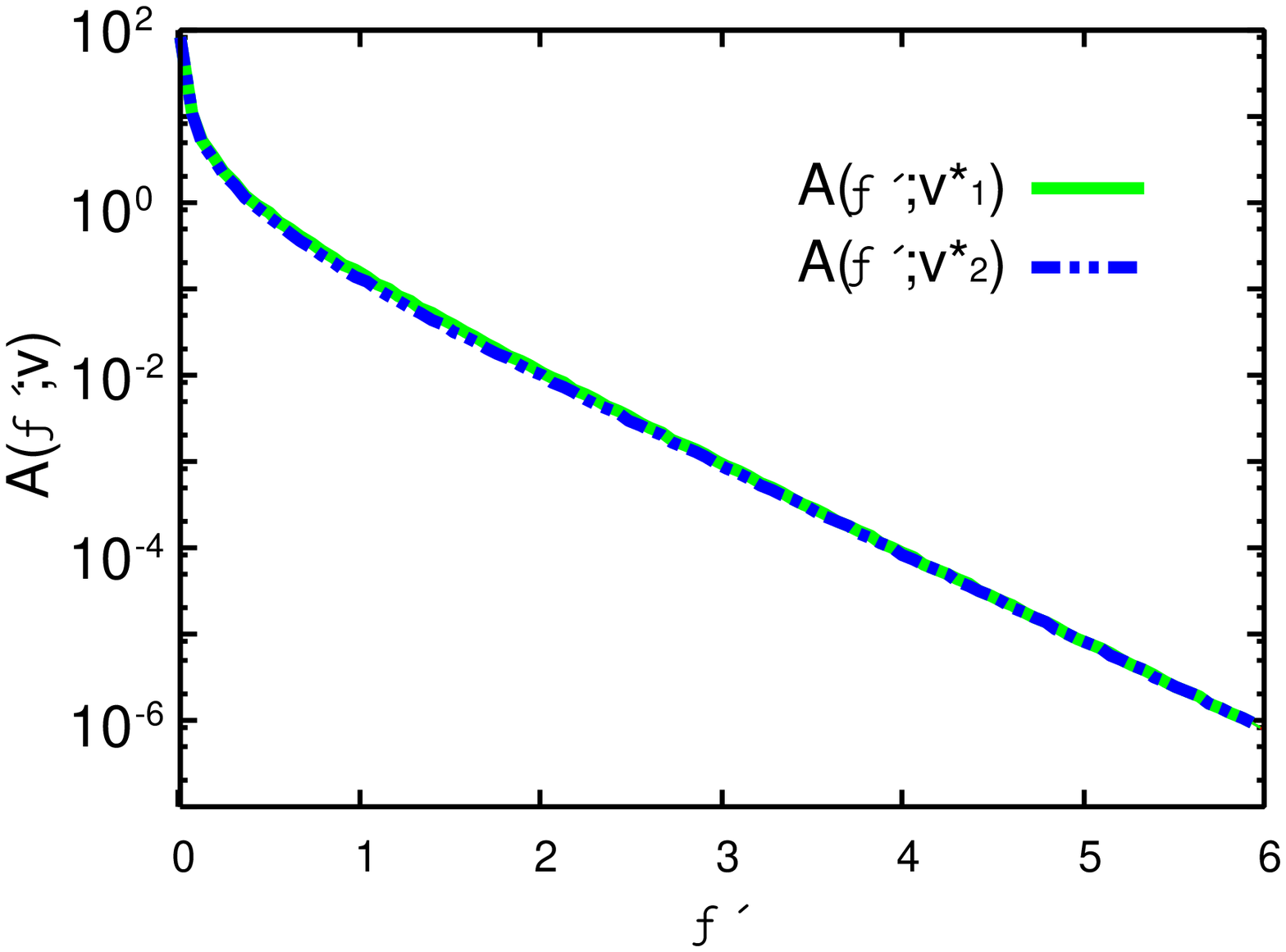}
      \caption{}
      \label{fig7}
    \end{center}
  \end{figure}

     \indent
     By increasing the Hill-coefficient $\beta$,
      $A(v)$ becomes much smaller than unity which asymptotically
      go to zero, even if the value of $u_{(2)}$ is the same level as $u_{(1)}$ or
      $K_v$ as is shown in Fig. \ref{fig7}.
      Note that the conditions $u_{(2)} < u_{(1)}$ and $u_{(2)}<K_v$ have to be satisfied.
      The value of the slope $A(v)$ shows an exponential decrease with $\beta$.
      Hence, $N_{(1)}$ is sustained at an almost constant level and the
     population size regulation of cell type 1 is realized with a
     sufficiently large $\beta$.

     \section{Model I\hspace{-.1em}I\hspace{-.1em}I:~
 Proportion Preservation of Two Cell Types}
  \indent\indent
  For precise body plan or for tissue homeostasis, proportion regulation
  of the number of each cell type is required. The fraction of each
  cell type has to be sustained at a certain range, against the change
  in the total number of cells.  Here we modify the kinetics of $v$ in
  the previous model \mbox{I\hspace{-0.1em}I} to seek for the
  possibility of the proportion regulation. With this modification, we
  will show that the population fraction of the two types of cells
  is kept at a certain level against the change of $N$.\\

      \indent
     Here, the kinetics of $v$ is modified as follows,
     \begin{align}
      \intertext{{\bf model I\hspace{-0.1em}I\hspace{-0.1em}I}}
      \frac{dv(t)}{dt} &= {\rm g_3}(u_1,\dots,u_N,v)\nonumber\\ 
      &= c_{v1}\sum_{i=1}^{N}{\frac{u^{\beta}_i(t)}{\tilde{K}^{\beta}_{v}+u^{\beta}_i(t)}}
      -c_{v2} v(t)\sum_{i'=1}^{N}{\frac{\tilde{K}^{\beta}_{v}}{\tilde{K}^{\beta}_{v}
      +u^{\beta}_{i'}(t)}}-v(t)\label{B2}
     \end{align}
     The modification to model \mbox{I\hspace{-0.1em}I} is just an
     addition of the  second term in Eq. (\ref{B2}). In other words,
     each cell in this model also contributes to the degradation of the
     signal $v$. \\

\begin{figure}[htb]
     \begin{center}
       \includegraphics[width=12cm]{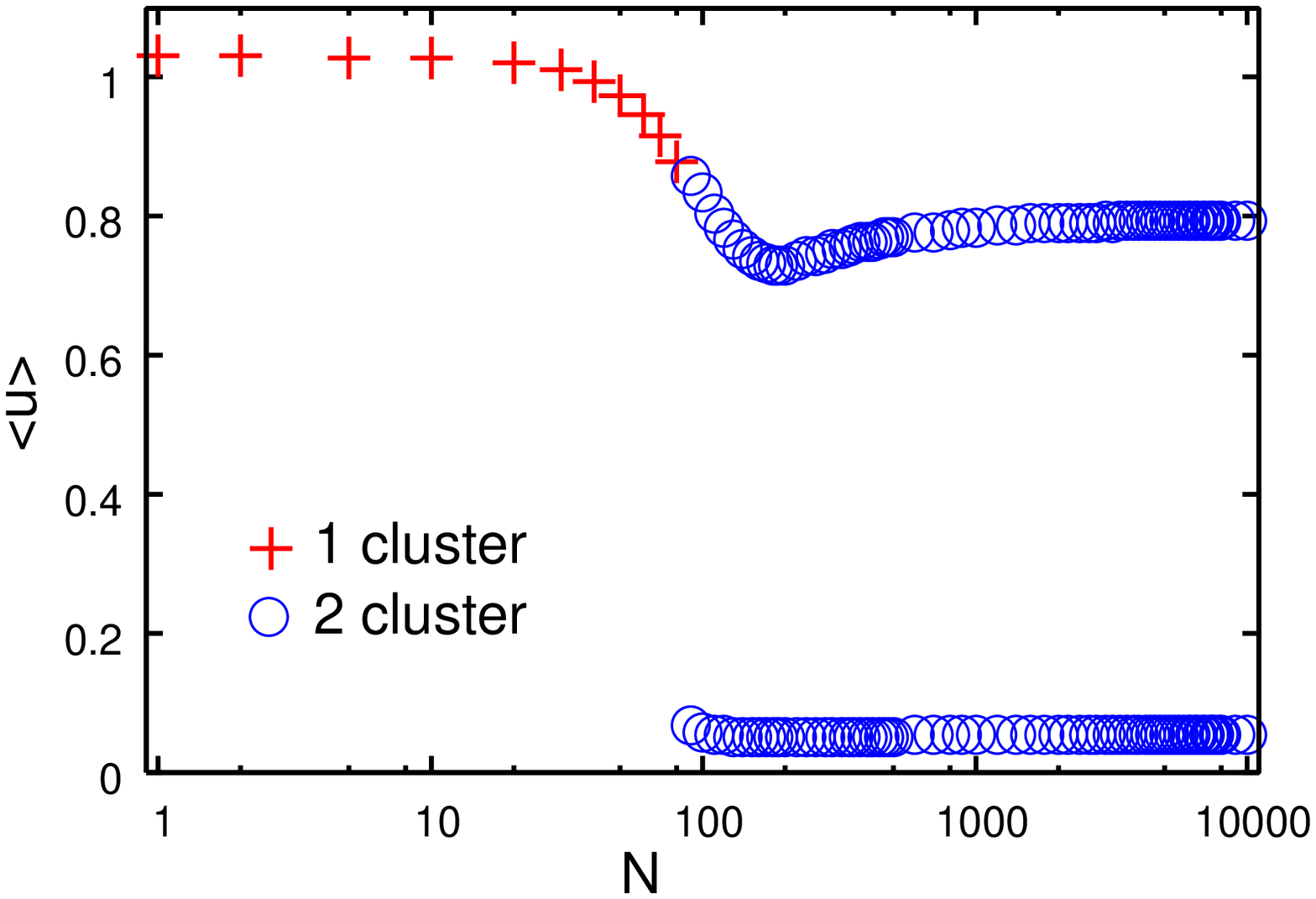}
      \caption{}
      \label{fig8}
    \end{center}
  \end{figure}

     \indent
     As in the previous model, the cellular states fall on stationary states,
     and the bifurcation of the stationary state from a single-cluster to two-cluster states are
     observed with the increase in $N$ (Fig. \ref{fig8}). Here, we
     first note that the two-cluster state remains stable over a wide
     range of $N$. 
     Indeed, non-zero $N_{(1)}$ exists so that
     $v_{1}^{*}<v(N_{(1)},N)<v_{2}^{*}$ is satisfied even for sufficiently large $N$. \\

\begin{figure}[htb]
    \begin{center}
      \includegraphics[width=12cm]{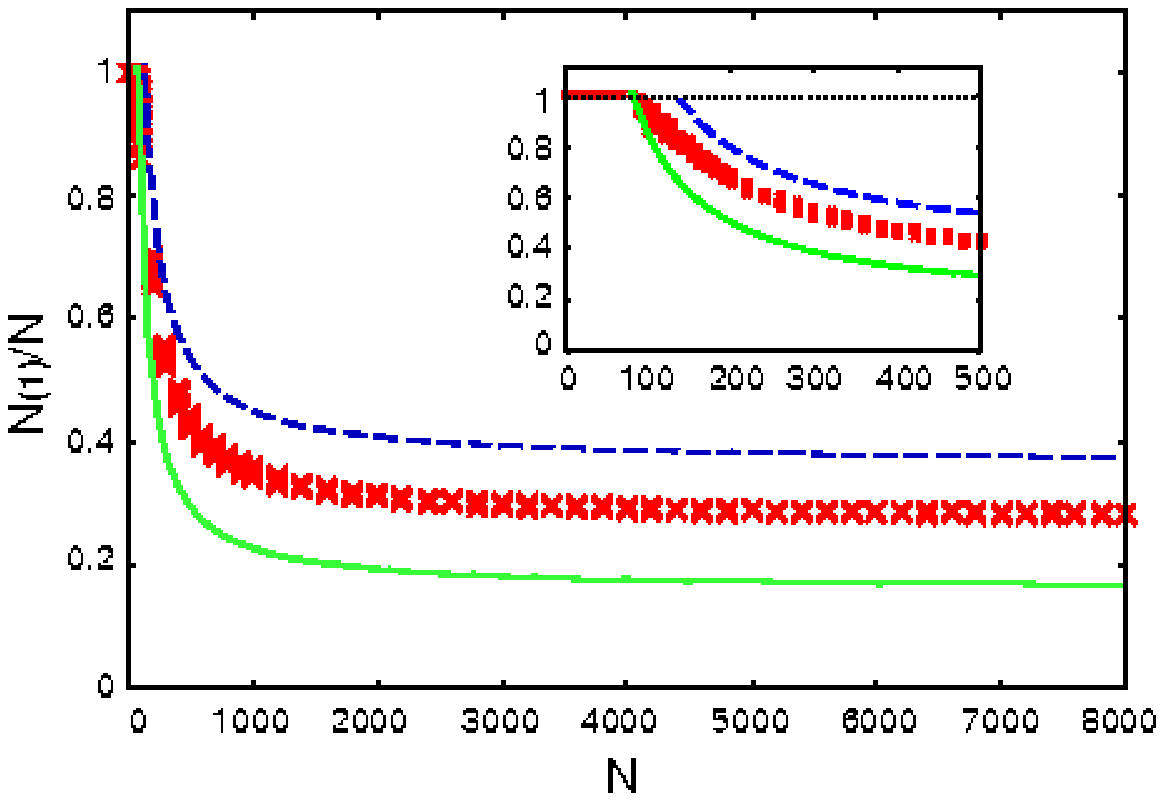}
      \caption{}
      \label{fig9}
     \end{center}
  \end{figure}

     \indent
     Next, we study the population distribution of two cell types. As shown
     in Fig. \ref{fig9}, the ratio $N_{(1)}/N$ stays at a constant level
     against the change of $N$. In the same way as in the previous section,
     the dependency of $N_{(1)}$ on $v$ and $N$ for a two-cluster state is
     written as,
     \begin{align}
      \frac{N_{(1)}(N,v)}{N}&=\tilde{A}(v)+\frac{\tilde{B}(v)}{N}
      \label{a14},
      \end{align}
      \begin{align}
       \tilde{A}(v)
      &= 
       \left[{\scriptstyle 
       1 + 
       \left( 
       \frac{ c_{v1}u_{(1)}^{\beta}(v) - c_{v2}\tilde{K}_{v}^{\beta}v }{
       c_{v2}\tilde{K}_{v}^{\beta}v - c_{v1}u_{(2)}^{\beta}(v)
       } 
       \right)
       \left(
       \frac{ \tilde{K}_{v}^{\beta}+u_{(2)}^{\beta}(v) }{
        \tilde{K}_{v}^{\beta}+u_{(1)}^{\beta}(v) }
       \right)
      }\right]^{-1}, \label{eq14}\\
      \tilde{B}(v)
      &= \frac{
       {\scriptstyle v}
      }{
       {\scriptstyle
      c_{v1} \left\{
      u_{(1)}^{\beta}(v)/(\tilde{K}_{v}^{\beta}+u_{(1)}^{\beta}(v) )
      -u_{(2)}^{\beta}(v)/(\tilde{K}_{v}^{\beta}+u_{(2)}^{\beta}(v))
      \right\}
      + c_{v2}\tilde{K}_{v}^{\beta} v \left\{
      1/(\tilde{K}_{v}^{\beta}+u_{(2)}^{\beta}(v) )
      -1/( \tilde{K}_{v}^{\beta}+u_{(1)}^{\beta}(v) )
      \right\}
      }}.
      \end{align}
      Here, $\tilde{B}(v)>0$ is always satisfied.
      Because $v$ satisfies $v^{*}_1< v < v^{*}_2$ 
      for the existence of a two-cluster state, 
      $N_{(1)}/N$ is within the range 
      $ ( {\scriptstyle \tilde{A}(v_1^*)+ \tilde{B}(v_1^*)/N}) 
      < \frac{N_{(1)}(N,v)}{N} <
      ( {\scriptstyle \tilde{A}(v_2^*)+\tilde{B}(v_2^*)/N} )$
      for each $N$.
      As a result, when $N$ is sufficiently large, the possible range of
      $N_{(1)}/N$ is given by
     \begin{align}
      \tilde{A}(v_{1}^{*})< \frac{N_{(1)}}{N}
      <\tilde{A}(v_{2}^{*}). \label{eq16}
     \end{align}
      From the above expression of $\tilde{A}(v)$, if the condition $(v_2^*/u_{(1)}^{\beta}(v_2^*))<
      c_{v1}/(c_{v2}K_v^{\beta})<(v_1^*/u_{(2)}^{\beta}(v_1^*))$
      is satisfied, $\tilde{A}(v)$ is within $0<\tilde{A}(v)<1$. 
      This is the case for the parameter values in Fig. \ref{fig9}.
      Thus, the cell type ratio of a two-cluster state has to be within the range
      given by Eq. (\ref{eq16}), so that its ratio is insensitive to the
      change of the total number of cells.
      In addition, by increasing the Hill-coefficient $\beta$,
      the range given by Eq. (\ref{eq16}) gets narrower.
      Thus, the ratio $N_{(1)}/N$ is more accurately regulated.
      As $\beta$ goes to infinity the range approaches its minimum, where the boundary is given by
      $\tilde{A}_{\infty}(v)=v/(c_{v1}/c_{v2}+v)$.\\

      \indent
      Note that $\tilde{A}(v)$ here is positive and is not necessarily
      small, in contrast to $A(v)$ in Eq. (\ref{a10}) for the model \mbox{I\hspace{-.1em}I}.
      Inclusion of the second term in Eq. (\ref{B2}) allows for this behavior, and the proportion regulation of cell types is achieved over a wide range of cells.

    \section{Cell Differentiation Model with Random Network}
    \indent\indent
    Here, we briefly discuss a general situation of cell differentiation models
    with  intracellular dynamics and intercellular interactions with more
    genes (chemical species).
    As an example, We use the cell differentiation models of
    Kaneko-Yomo or Furusawa-Kaneko~\citep{Kaneko97,Kaneko99,Furusawa98,Furusawa01}. Here, we aim at
    demonstrating that the regulative behavior of cell differentiation
    in the previous sections generally works, which at the
    same time may provide a possible 
    explanation for differentiation phenomena observed in their models.
    For the following analysis, we use one of the models (FK model) introduced in~\citep{Furusawa01}, while it is straightforwardly extended to other
    models.\\

    \indent
    In the FK model each cell has 
    intracellular metabolic dynamics, and grows by uptake of the
    nutrients in the medium, and divides when the abundances of
    chemicals in the cell goes beyond some threshold. Accordingly, the
    total cell number $N$ is
    also a time-dependent variable. As the cells share the same
    medium, they interact with other cells through uptake from the medium and
    exchange of chemicals with it.\\

    \indent
    The state of cell $l$ is expressed by $P$ different metabolites, 
    $\vec{x}^{(l)}=(x^{(l)}_1,\ldots,x^{(l)}_P)^{T}$, and the nutrients are
    $\vec{X}=(X_1,\ldots,X_{Q})^T$, $(Q \leq P)$. 
    The dynamics of the $i$-th metabolite in cell $l$ is given as
    follows, 
    \begin{align}
     \frac{dx^{(l)}_i(t)}{dt} =
     F_i \left( \{ x^{(l)}_i(t)\}, X_i(t);
     \{C_{ijk}\},\{{\sigma}_{i}\} \right),\label{a18}
    \end{align}
    A change in the concentration of the $i$-th nutrient in the medium with the
    volume $V$ is given by,
    \begin{align}
     \frac{dX_i(t)}{dt} &=
     D_{\rm env}(S_i-X_i(t))
     -\frac{D}{V}\sum_{m=1}^{N}{(X_i(t)-x^{(m)}_i(t))}.
     \label{a19}
    \end{align}
    $S_i$ is the external source of the nutrient, $D_{\rm env}$ is the
    diffusion constant between the nutrient reservoir and the medium,
    and $D$ is that across the cell membrane.
    Each cell grows through uptake of nutrients and
    changing them to other metabolites by Eq. (\ref{a18}).
    As the cells share the same medium, they interact with each other through competition for nutrients.
    Here we confine our consideration only to 
    the behavior of nutrients $\{X_i\}$ in the stationary states
    for fixed $N$, and to obtain the behavior of the stationary states
    as a function of $N$.\\

    \indent
    Because the stationary state satisfies the condition
    $dx^{(l)}_i/dt=0$ and $dX_i/dt=0$,
    \begin{align}
     F_i \left( \{ x^{(l)}_i\}, X_i;
     \{C_{ijk}\},\{{\sigma}_{i}\} \right)=0,\label{a20}\\
     D_{\rm env}(S_i-X_i)
     -\frac{D}{V}\sum_{m=1}^{N}{(X_i-x^{(m)}_i)} = 0.\label{a21}
    \end{align}
    From Eq. (\ref{a20}), possible stationary states of each cell, i.e.,
    stationary solutions of $\vec{x}^{(l)}$, are obtained as a function of $\vec{X}$. 
    Next, we describe how $\vec{X}$ varies with $N$.
    As in the previous sections, 
    we assume that the cell population takes an $M$-cluster
    state in the stationary state for a given $N$.
    By solving Eq. (\ref{a21}) for $X_{i}$, one obtains 
    \begin{align}
      X_{i} =\frac{ \sum_{k=1}^{M}{ R_{k}{\hat{x}}_{i}^{(k)} }
        +\left(\frac{ VD_{\rm env} }{ D }\right) \frac{S_{i}}{N} }
      { 1+ \left(\frac{ VD_{\rm env} }{ D }\right) \frac{1}{N} },\label{a22}
    \end{align}
    where ${\hat{x}}_{i}^{(k)}$ is the cell type $k$ in an $M$-cluster state,
    and $R_{k} = N_{k}/N$, with $N_{k}$ as the number of type 
    $k$ cells in the population.
    $X_i$ is represented as a function of $N$ and $\{R_k\}$.
    The stability condition of the $M$-cluster state of concern 
    is expressed by $N$ and $\{R_k\}$,
    from Eq. (\ref{a20}) and Eq. (\ref{a22}).
    Thus, the realization of an $M$-cluster state
    depends on the number of cells or the ratio of cell types.  Regulation of each cell type, 
    as observed in \citep{Kaneko97,Kaneko99,Furusawa98,Furusawa01}, is expected accordingly.

    \section{Summary and Discussion}
    \indent\indent
    Through the analysis of several models, we see,
     i) a switch of cell types via an increase of the total cell number,
     and ii) diversification to two cell types.
     In addition, when the cells differentiate 
     to two types,
     size preservation of a specific cell type or
     proportion preservation of two cell types appears,
     depending on the interaction form with other cells.
     These behaviors are explained
     as a bifurcation of cell states via the intercellular interactions.
     First, possible cell types $u_{(1)}$ and $u_{(2)}$ are generated by
     a single positive feedback loop, which works as a module for bistability.
     Secondly, intercellular signal $v$ works as a bifurcation parameter,
     whose abundances determine the actual cell types.
     This bifurcation parameter is a function of
     the number of each cell type, depending on the intercellular interactions.
     Then, the resulting bifurcation parameter has to be determined self-consistently.
     This constraint restricts the number distribution of the cell types, 
     which gives the mechanism of the regulation of the cell differentiation.\\

     \indent
     In model I, because the total cell number simply corresponds to
     the bifurcation parameter of cell states, the switch of the cell types
     by the total cell number is straightforward.
     In model I\hspace{-.1em}I and I\hspace{-.1em}I\hspace{-.1em}I,
     since intercellular couplings change the bifurcation parameter,
     the transition from the single-cluster state of $u_{(1)}$ to a two-cluster state occurs
     by the increase in the total cell number.
     In model I\hspace{-.1em}I, the cell-type 2 contributes only weakly 
     to the increase of $v$, compared with the cell-type 1.
     Thus, the amount of $v$ mainly depends on the number of the cell-type 1.
     In contrast, in model I\hspace{-.1em}I\hspace{-.1em}I, the cell-type 2 degrades $v$.
     As a result, the amount of $v$ depends on the number ratio of two cell-types.\\

     \indent
     If a gene expression network shows bistability with a bifurcation structure as in Fig. 1,
     cell differentiation is a general consequence when cell-cell couplings are introduced.
      An important point here is that the same intracellular module can be used
      in several different biological contexts by modifying only the
      intercellular interaction.
      This is quite useful in an evolutionary perspective because
      new biological functions can be added by incorporating new
      interactions while preserving the intracellular core module.\\

      \indent
      Here, we discuss several biological examples that may correspond to our models.
      First, we refer to the cell cycle machinery in {\it Xenopus},
      where a Cdc2 positive feedback loop makes a bifurcation with regards to 
      the amount of cyclin B, and introduces bistability as in Fig. 1.
      In this system, an increase in the DNA amount in the early embryo 
      induces the transition from the low Tyr15 phosphorylation state of Cdc2 to the high Tyr15
      phosphorylation state, which seems to cause the mid-blastula transition in {\it Xenopus}
      ~\citep{Novak93,Hartley96,Sha03}.
      The induced differentiation may correspond to that observed in model I.\\

      \indent
      Secondly, as for model  I\hspace{-.1em}I,
      consider the maintenance of the hematopoietic stem cells in mammals, where
      osteoblasts work as a stem cell niche~\citep{Calvi06}.
      The stem cells compete for some chemical factor representing this niche, and the cells which cannot take the factor
      differentiate to specific hematopoietic lineages.
      It has been discussed that the regulation of the stem cell population size is realized
      through the competition for the factor,
      to which responsibility decreases through the differentiation process~\citep{Radtke04, Adams06}.
      Indeed, it is observed that the expression of Notch1, which is a candidate
      for the involvement in the niche-stem interaction,
      disappears after commitment to the lymphoid lineage~\citep{Radtke04}.
      The differentiation in the hematopoietic system may
      correspond to that studied in model I\hspace{-.1em}I.\\

      \indent
      Thirdly, an example for model  I\hspace{-.1em}I\hspace{-.1em}I
      is given by the proportion regulation of prestalk-cell types
      and prespore-cell types in the {\it Dictyostelium} slug.
      Differentiation to prespore cells is induced by cAMP, and
      the cell state is maintained by a positive-feedback loop of
      prespore cell specific adenylyl cyclase G activity
      ~\citep{Hopper93,Williams06, Alvarez-Curto07}.
      On the other hand, differentiation-inducing factor-1 (DIF-1)
      is necessary for the differentiation from
      a prespore-cell to a prestalk-cell
      (at least for the differentiation to pstO 
      which is a subtype of the prestalk-cell)~\citep{Williams06,Kay01}.
      As an intercellular interaction, 
      this DIF-1 is produced by prespore-cells, and 
      are degraded by prestalk-cells.
      This cell-type specific induction/destruction of DIF-1
      is responsible for the proportion preservation
      as studied in model I\hspace{-.1em}I\hspace{-.1em}I.\\

      \indent
      Although we confine our analysis to a system with
      only fixed point solutions, oscillatory and other dynamical behaviors are
      often observed in biological systems.  The analysis we introduced here is also applicable
      to such cases, as long as there are bifurcations of attractors with the change in relevant
      chemical concentrations which are influenced by cell-cell interactions.  On the other hand,
      oscillatory behaviors may bring about richer bifurcations, as well as clustering of cells
      with regards to the oscillation phase or
      amplitude, as has been discussed in models with intra-cellular oscillatory dynamics
      and cell-cell interactions~\citep{Kaneko94,Kaneko97,Koseska07,Ullner07}.
      The study of possible forms on differentiations and regulations in such dynamical systems will
      be important in future.
      In multicellular systems, cells behave in coordination
      by taking advantage of communication with other cells.  Such collective behavior
      is a result of interacting systems with intra-cellular gene expression dynamics.
      The present self-consistent determination of bifurcation parameters through cell-cell interactions will
       be essential to understand organization in multicellularity.

\section*{Acknowledgements}
The authors would like to thank M. Tachikawa, N. Kataoka,
K. Fujimoto, and S. Ishihara for stimulating discussions.

\renewcommand{\refname}{References}
\bibliographystyle{elsart-harv}
\bibliography{differentiation_qbio}

\newpage
\section*{Captions}
{\bf Figure 1:}\\
\indent
The value $u$ of the fixed point solution as a function of the signal  concentration $v$ in our model.
Solid line indicates the stable solution, while the dotted line indicates the unstable one.\\

\noindent
{\bf Figure 2:}\\
\indent
The stationary states of $u_i$ in model I
are plotted against the total cell number $N$. At the interval $N^{*}_{1} \leq N \leq N^{*}_{2}$,
two different cell states coexist.
The parameter value $c_{1}$ is set at 0.005.\\

\noindent
{\bf Figure 3:}\\
\indent
The ratio of the number of each cell type ($\times$ for $N_{(1)}$ and $\square$ for $N_{(2)}$) plotted against the total cell number $N$, for model I.
The initial values of $u_i$ are chosen randomly from the interval
of $u_i \in [0,1]$. The parameter value is $c_{1}= 0.005$.\\

\noindent
{\bf Figure 4:}\\
\indent
The fixed point values of $u_i$ in model \mbox{I\hspace{-0.1em}I} are plotted against the total cell number $N$.
At each $N$, 100 initial conditions are chosen.  The expression levels of
$u_i$ for a single cluster ($+$) and two-cluster solutions ($\circ$) are plotted as a function of $N$. The value for two-cluster solutions is the average over initial conditions.  The parameter values are set at $K_{v}=2.0$, ${\beta}=2.0$, $c_{2}=0.1$.\\

\noindent
{\bf Figure 5:}\\
\indent
The stationary state of a single-cluster solution for model \mbox{I\hspace{-0.1em}I}.
Solid line indicates $u_i$ of the stable fixed solution, while
the broken line denotes that of the unstable one.
The parameters are $K_{v}=2.0$, ${\beta}=2.0$, $c_{2}=0.1$.\\

\noindent
{\bf Figure 6:}\\
\indent
The number of cell type 1 ($\times$) is plotted against the total cell number 
in model \mbox{I\hspace{-0.1em}I}.
The initial condition of $u_i$ is chosen randomly from the interval $u_i \in [0,1]$.
Solid and broken lines indicate $N_{(1)}(N,v_1^*) = -A(v^{*}_1)N + B(v^{*}_1)$
and $N_{(1)}(N,v_2^*) = -A(v^{*}_2)N + B(v^{*}_2)$, respectively, where
$A(v^{*}_1) = 0.011$,
$B(v^{*}_1) = 24$,
$A(v^{*}_2) = 0.0099$, and
$B(v^{*}_2) = 93$.
The parameters are $K_{v}=2.0$, ${\beta}=2.0$, $c_{2}=0.1$.\\

\noindent
{\bf Figure 7:}\\
 \indent
 The slope $A(\beta;v)$ is plotted as a function of $\beta$.
 Here, $A(\beta;v)$ for two different values of $v$, i.e., $v_1^*$
 and $v_2^*$ are plotted, which agree within the resolution of the plot in the figure.
 The parameter values are $K_{v}=2.0$, $c_{2}=0.1$.\\

\noindent
{\bf Figure 8:}\\
\indent
The fixed point solutions of model \mbox{I\hspace{-0.1em}I\hspace{-0.1em}I} plotted against the total cell number $N$.
At each $N$, 100 initial conditions are chosen.  The expression levels of
$u_i$ for a single cluster ($+$) and two-cluster solutions ($\circ$) are plotted as a function of $N$.
The value for two-cluster solutions is the average over initial conditions.  
The parameter values are
$\tilde{K}_{v}=0.2$, ${\beta}=2.0$, $c_{v1}=c_{v2}=0.005$.\\

\noindent
{\bf Figure 9:}\\
\indent
The ratio of the number cell type 1 $N_{(1)}$ to the total cell number $N$ is plotted against $N$ for model \mbox{I\hspace{-.1em}I\hspace{-.1em}I}.
The initial condition of $u_i$ is chosen randomly from the interval
of $u_i \in [0,1]$.
Solid and broken lines indicate
$N_{(1)}(N,v_1^*)/N = \tilde{A}(v^{*}_1) + \tilde{B}(v^{*}_1)/N$
and
$N_{(1)}(N,v_2^*)/N = \tilde{A}(v^{*}_2) + \tilde{B}(v^{*}_2)/N$,
respectively, where
$\tilde{A}(v^{*}_1) = 0.16$,
$\tilde{B}(v^{*}_1) = 69$, 
$\tilde{A}(v^{*}_2) = 0.36$, and 
$\tilde{B}(v^{*}_2) = 86$.
The parameter values are
$\tilde{K}_{v}=0.2$, ${\beta}=2.0$, $c_{v1}=c_{v2}=0.005$.

\newpage

\end{document}